# Superconducting proximity effect in epitaxial Nb(110)/Au(111)/Nb(110) trilayers

Hiroki Yamazaki[1, *], Nic Shannon[2, 3] and Hidenori Takagi[4, 5]

[1] *RIKEN, Wako, Saitama 351-0198, Japan*
[2] *Okinawa Institute of Science and Technology Graduate University, Onna-son, Okinawa 904-0495, Japan*
[3] *H.H. Wills Physics Laboratory, Tyndall Avenue, University of Bristol, BS8 1TL, UK*
[4] *Department of Physics, University of Tokyo, Bunkyo, Tokyo 113-8656, Japan*
[5] *Max-Planck-Institute for solid state research, Heisenbergstrasse 1, 70569 Stuttgart, Germany*

[*]yamazaki@postman.riken.jp

Single-domain, epitaxially-grown Nb(110)/Au(111)/Nb(110) trilayers are studied in order to elucidate the superconducting proximity effect in the Au(111) layer. The critical current density, superconducting coherence length, and superconducting transition temperature, all show a non-monotonic dependence on the thickness of the Au layer. The analysis of the experimental data reasonably argues that a form of order-parameter oscillation is intrinsic to the proximity-induced Cooper pairs in the Au layer. We find that relative phase of the superconducting order parameter in the two Nb(110) layers changes sign as a function of Au(111) layer thickness $t_{Au}$, offering a route to achieving a π junction for 0.94 nm<$t_{Au}$<1.88 nm.

# I. INTRODUCTION

Superconducting electronics, in which currents are carried by superconducting pairs of electrons, are used in existing detector technologies, and have great potential for future application in quantum computing. An essential element for superconducting circuits is the π junction, in which the phase of a superconducting wave function is inverted. One promising approach to building such a junction is to exploit the interplay between superconductivity and magnetism in superconductor/ferromagnet/superconductor (SC/FM/SC) heterostructures [1]. Such multilayer structures are widely believed to achieve a Fulde-Ferrell-Larkin-Ovchinnikov (FFLO) state [2,3], in which the superconducting proximity effect drives a superconducting order with oscillating phase within the FM layer [4]. Mesoscopic superconductor/normal metal/superconductor (SC/NM/SC) structures, in which Andreev levels in the normal metal are driven out of equilibrium by the application of a voltage, have also been argued support a π junction [5]. None the less, both of these approaches suffer from the serious drawback that ferromagnetic elements, and non-equilibrium states, tend to suppress superconductivity, forcing devices to operate at lower temperatures, and with lower critical current densities. It is therefore highly desirable to find a route to constructing a π junction which does not rely on introducing FM elements, or driving the system out of equilibrium.

In this article we show that, in epitaxial Nb(110)/Au(111)/Nb(110) trilayers, the equilibrium superconducting state presents strong evidence of 0-π state transitions as a function of the Au-layer thickness, without the need for ferromagnetic elements, or a bias voltage. The analysis of the experimental data reasonably argues that a form of order-parameter oscillation, similar to the FFLO-like state in FM for SC/FM/SC junctions, is intrinsic to the proximity-induced Cooper pairs in the Au(111) layer. We tentatively ascribe this effect to the spin-orbit coupling (SOC) within the Au layer. Where SOC lifts the degeneracy between "up" and "down" spin electrons, electron pairs can form with a non-zero momentum. This leads to FFLO-like oscillations of the superconducting order parameter in real space. It should be noted that, in face-centered-cubic (fcc) lattice like that of Au, the inversion symmetry is broken in the direction perpendicular to the (111) plane because of the ABCABC… stacking sequence of atomic planes, where a lack of inversion symmetry leads to the emergence of SOC. For Au, moreover, a prominent contribution of the 5$d$-orbital to conduction electrons has been revealed experimentally: the 5$d$ weight to the total density of states at the Fermi level was estimated to be 40-60% [6]. The 5*d* rather than 6*s* character will result in a considerable enhancement of SOC for conduction electrons. In fact, the SOC strength of conduction electrons, which was determined by the magneto-optical Kerr effect, was 120 meV in Au [7]. This value is close to an energy splitting of 110 meV due to Rashba SOC for the *sp*-derived *surface* state of Au(111) [8,9].

We have previously shown that both the trilayers of Nb(110)/Au(111)/Fe(110) [10] and Nb(110)/Au(111)/Co(0001) [11] exhibit long-period oscillations in the superconducting transition temperature $T_c$ as a function of the Au(111) layer thickness. The oscillation periods are the same (~2.1 nm) irrespective of the difference in FM between Fe and Co, and, more importantly, the oscillations occur for the Au layer thicknesses beyond the range of the Ruderman-Kittel-Kasuya-Yosida (RKKY) coupling [11]. These features suggest that it could be possible to achieve superconducting order-parameter oscillations in the Au(111) layer by the proximity to a Nb. Here we show how this can be accomplished in epitaxial Nb(110)/Au(111)/Nb(110) trilayers, paving the way for the construction of a superconducting π junction.

## II. SAMPLE PREPARATION

The starting point for our study was a series of Nb(110)/Au(111)/Nb(110) trilayers prepared on single crystals of $Al_2O_3(11\bar{2}0)$ using molecular-beam-epitaxy (MBE). A schematic diagram of the sample structure is shown in Fig. 1. Details of the preparation, structural characterization and measurements are essentially the same as those for Nb(110)/Au(111)/Fe(110) [10] and Nb(110)/Au(111)/Co(0001) [11]. First, a Nb layer of 20.0 nm(=$t_{Nb}$) was grown on the substrate at $T_s$=400ºC ($T_s$: the substrate temperature). The condition $T_s$<450ºC keeps the oxygen of the $Al_2O_3$ substrate from diffusing into the Nb layer [12]. Subsequently, a Au layer with a thickness of $t_{Au}$=0.17–4.00 nm and a Nb layer with a thickness of $t_{Nb}$ were deposited at $T_s$=200ºC. The trilayer was finally capped with a Au layer of 1.74 nm at $T_s$=100ºC in order to avoid oxidization. One atomic monolayer (1 ML) of Au(111) corresponds to $t_{Au}$=0.2355 nm.

## III. RESULTS AND ANALYSIS

During the sample growth, reflection high-energy electron diffraction (RHEED) patterns were measured for the surface of each layer. The patterns in the left and right columns of Fig.2 are typical of the samples for 0.40≤$t_{Au}$≤2.10 nm and $t_{Au}$≥2.25 nm, respectively. The Au-layer surface presented fine streak pattern for $t_{Au} \geq 0.40$ nm, indicating that epitaxial layer-by-layer growth occurs from the early stages of the Au-layer growth. For the upper Nb layer, the RHEED pattern changed from that of single domain ($0 < t_{Au} \leq 2.10$ nm) to that of twinned domain ($t_{Au} \geq 2.25$ nm) structure with a narrow interval (2.10-2.25 nm) of transition. This structural change is due to the lattice mismatch between Au and Nb, and the structure of the upper Nb layer depends entirely upon $t_{Au}$. The twinning is accompanied by a ±120° rotation of the in-plane <001> axis around the <110> axis that is perpendicular to the surface. Off-axial X-ray diffraction measurements support this result. For the samples of $t_{Au} > 2.10$ nm, therefore, it is due to the presence of the twinned domains (i.e., a form of structural disorder) in the upper Nb layer that we *cannot* see a systematic change of $T_c$ ($T_c^{resistive}$ and $T_c^{magnetic}$) as a function of $t_{Au}$ (see Fig. 3). Here $T_c^{resistive}$ is defined as the temperature at 50% of the residual resistivity, and $T_c^{magnetic}$ is as the onset point of the diamagnetic transition. Typical temperature dependences of normalized resistivity $R_n$ ($H=0$ Oe) and of normalized magnetic susceptibility $\chi_n$ ($H=0.1$ Oe, $H \perp$ surface) are shown in Fig. 4. In the article below we will focus on the single-domain samples ($0 < t_{Au} \leq 2.10$ nm) whose superconducting properties are uniquely determined by $t_{Au}$.

The average electron mean free paths $l^{Au}=140\pm40$ nm and $l^{Nb}=3\pm1$ nm were calculated from the residual resistivity as a function of $t_{Au}$ by performing a theoretical fit using a parallel register model. We also estimated the superconducting coherence lengths $\xi_{//}$ and $\xi_{\perp}$ from the measurements of the upper critical fields $H_{c2\perp}$ and $H_{c2//}$ using the expressions: $\xi_{//}=(\varphi_0/2\pi H_{c2\perp})^{1/2}$ and $\xi_{\perp}=(\varphi_0/2\pi)(\xi_{//}H_{c2//})^{-1}$, where $\varphi_0$ is the flux quantum and the directions, $//$ and $\perp$, are referred to with respect to the surface. The superconducting coherence length of the Nb layer at zero temperature was found to be $\xi_{//}^{Nb}(0) \sim \xi_{\perp}^{Nb}(0) \sim 20$ nm by means of an extrapolation of the values, $\xi_{//}^{Nb}(T)$ and $\xi_{\perp}^{Nb}(T)$ for $T=0.3T_c$–$0.9T_c$. These results imply that charge transport through the Au spacer layer is ballistic ($l^{Au} >> t_{Au}$) and the superconductivity of the Nb layers is in the dirty limit ($\xi^{Nb}(0) > l^{Nb}$). We also point out that the sample total thickness, $t_{total}=20.0+t_{Au}+20.0+1.74$ nm (excluding the substrate), is comparable to the field-penetration depth of $\lambda(0) \sim 39$ nm in bulk Nb (ref. [13]).

The next step is to estimate the superconducting critical current density $J_c$ in zero magnetic field. This was estimated from the width ($\Delta M$) of the *M-H* hysteresis curve for *H parallel* to the sample plane, using the result $\Delta M = J_c t /20$ derived for a Bean model of a long slab of thickness $t$ [14]. Here, $J_c$ is measured in A/cm$^2$, $t_{total}$ is taken for $t$ [cm], and $\Delta M$ [gauss] is measured at 0 Oe. Our results for $J_c$ are shown in Fig. 5(a), together with a typical *M-H* curve (inset) in a sample with $t_{Au}$=1.13 nm (~5 ML). The vertical error bars on $J_c$ reflect the fact that the magnetization decays exponentially to a saturation value in a time of order $10^2$ sec when the system is held at 0 Oe, with a ~1% (~10%) reduction at 4.2 K (6.0 K). For samples with $t_{Au}$>0.5 nm (~2 ML), over a range of temperatures ~3.5 K<$T$<$T_c$, the temperature dependence of $J_c$ is well-described by the expression $J_c(T)=J_c(0)(1-T/T_c)^\alpha$, with α=1.4±0.1. This is in good agreement with the exponent α=1.39 obtained for critical current at 0 Oe in an 120 nm thick Nb film [15]. Ginzburg-Landau theory predicts α=3/2 near $T_c$ [16]. The values of $J_c$=(0.1–10)×$10^8$ A/cm$^2$ seen in Fig. 5(a) are extremely high, relative to the SC/FM/SC systems studied so far [1]. In fact, these $J_c$ values are close to the depairing current density $J_{dp}$ observed in Nb thin films using transport measurements [16]. As theoretical studies have shown that, in parallel fields, the $J_c$ in films with thickness ≤ λ can be comparable to $J_{dp}$ [17-19]. Especially in our samples, for $|H|$→0 ($H$ // film plane), vortices are trapped in and along the Au layer between the Nb layers, because the Au layer is placed at the middle of the trilayer and behaves itself as pinning centers.

The proximity effect in these Nb(110)/Au(111)/Nb(110) trilayers also displays a number of features. As shown in Fig. 5, neither $J_c$, $\xi_\perp$, nor $T_c$ are smooth, monotonic functions of the thickness $t_{Au}$. In particular, there are sharp features at $t_{Au}$=0.94 nm (~4 ML) and 1.88 nm (~8 ML). We wish to emphasize that, at these thicknesses, there are *no* changes in crystal structure. As mentioned in the previous section, crystal twinning actually occurs in the upper Nb layer, but only for $t_{Au}$>2.10 nm. The observed modulation in $J_c(t_{Au})$ and $T_c(t_{Au})$ is therefore intrinsic to Nb(110)/Au(111)/Nb(110) trilayers. Such systematic, non-monotonic variations in $J_c$ and $T_c$ as functions of the NM-layer thickness have not been reported before, and we infer that epitaxially grown multilayer systems of equivalently high quality are necessary for it to be observed.

Moreover, while the absolute value of $J_c$ changes as a function of temperature, the modulation of $J_c$ as a function of $t_{Au}$ persists, as shown in Fig. 5(a). For 0<$t_{Au}$≤0.94 nm(~4 ML), we see a monotonic decrease of $J_c$ as $t_{Au}$ increases; while for 0.94 nm≤$t_{Au}$≤1.88 nm(~8 ML) a convex function of $J_c(t_{Au})$ is recognized. This sort of oscillatory behaviour in $J_c(t_{Au})$ and its temperature dependence are quite similar, for example, to the behaviour of $I_c(t_{Ni})$ observed for Nb/Cu/Ni/Cu/Nb junctions (see Fig. 5 in Ref. [20]), where $I_c$ is the superconducting critical current in the junctions and $t_{Ni}$ is the Ni-layer thickness. These junctions have been confirmed to exhibit the 0-π state transitions as a function of $t_{Ni}$ [20]. An oscillatory behavior of $I_c(t_{FM})$, with accompanying 0-π state transition, has been predicted theoretically to occur in SC/FM/SC systems, where the SC order parameter in the FM layer undergoes FFLO oscillations [1]. We now explore how well the results of this theory fit the present case. For the clean, thin and strong FM layers, $I_c$ should vary with $t_{FM}$ *near* $T_c$ as (referred to Eq. (1) in Ref. [20] and Eq. (56) in Ref. [1]):

$$I_c = I_c^0 |\sin y| / y \qquad y = 2\pi E_{ex} t_{FM} / h v_f, \qquad (1)$$

where $E_{ex}$ is the exchange energy of FM, $h$ is Planck's constant, and $v_f$ is the Fermi velocity of FM. In applying this result to our case, we identify $I_c$ with $J_c$, $t_{FM}$ with $t_{Au}$, take the bulk value of $v_f^{Au}=1.39\times10^6$ m/sec (Ref. [21]), and treat $E_{ex}$ as an effective parameter characterizing the order parameter oscillations in the Au layer, i.e.

$$J_c = J_c^0 |\sin y| / y \qquad y = 2\pi E_{ex}^{eff} t_{Au} / h v_f^{Au}. \qquad (2)$$

This allows two fitting parameters, the overall scale $J_c^0$, and $E_{ex}^{eff}$. The resulting fits for $E_{ex}^{eff}=84.6$ meV and $J_c^0(4.2\text{ K})/J_c^0(6.0\text{ K})=5.0$ are shown in Fig. 6. Here the ratio $J_c^0(4.2\text{ K})/J_c^0(6.0\text{ K})=5.0$ corresponds to the temperature dependence $J_c^0(T)=J_c^0(0)(1-T/T_c)^\alpha$ for $\alpha\sim1.4$. Except for small $t_{Au}$, we see a reasonable agreement between Eq. (2) and experimental data, especially at 6.0 K (near $T_c$). The discrepancy particularly for $t_{Au}\leq2$ ML can be attributed to the interface disorder occurred in the region of ~2 ML thickness between the lower-Nb and the Au layer. This is supported by the results of the RHEED measurements: no fine streak pattern in RHEED was observed until $t_{Au}$ increases to 0.4 nm (~1.7 ML), suggesting a probable reduction in $l^{Au}$ and the presence of short-cuts between the Nb layers for $t_{Au}\leq2$ ML. From the picture of a pseudomagnetic effect in the Au layer, we can thus obtain a quantitative understanding of the oscillatory behavior in $J_c(t_{Au})$, and the singularity at $t_{Au}\sim4$ ML (8 ML) can be explained in terms of the 0→π (π→0) state transition.

The above picture is also sustained by the behavior of $T_c$ and $\xi_\perp$ as functions of $t_{Au}$. Firstly, at the thicknesses of $t_{Au}\sim4$ ML(~1 nm) and 8 ML(~2 nm), both the data of $T_c^{resistive}$ and $T_c^{magnetic}$ contain pronounced local minima as seen in Fig. 5(c). In the SC/FM/SC system, generally, the superconducting transition temperature as a function of $t_{FM}$ behaves differently obeying $T_c^{*0}(t_{FM})$ or $T_c^{*\pi}(t_{FM})$ according to the phase difference between the SC layers, where $T_c^{*0}(t_{FM})$ is for the 0 coupling and $T_c^{*\pi}(t_{FM})$ for the π coupling. We will observe higher $T_c^*$ ($T_c^{*0}(t_{FM})$ or $T_c^{*\pi}(t_{FM})$) as $T_c$, and the 0-π state transitions take place where the two curves cross [1, 22]. The crossing points therefore show local minima or kinks in $T_c(t_{FM})$, like those found in $T_c(t_{Au})$. In fact, the periodic change in $T_c(t_{Au})$ is similarly comparable to that of $T_c(t_{Ni})$ presented typically in (Ni/)Nb/Ni/Nb(/Ni) multilayers (see Fig. 3 in Ref. [22]). Secondly, the 0-π state transitions explain the way the perpendicular coherence length $\xi_\perp$ changes with $t_{Au}$ (see Fig. 5(b)). For $0<t_{Au}\leq4$ ML, where two Nb layers are in phase, the coherence length $\xi_\perp$ exceeds the thickness of a single Nb layer ($t_{Nb}$); then $\xi_\perp$ becomes comparable with the sample total thickness ($t_{total}$), when approaching $T_c$ (a 'normal' superconducting proximity effect). However for $4\leq t_{Au}\leq8$ ML, the coherence length $\xi_\perp$ is locked to $t_{Nb}$, and only exceeds this again for $t_{Au}>8$ ML. The suppression of the measured value of $\xi_\perp$ can be attributed to the $\pi$ phase difference between the two Nb layers, i.e. a nodal plane of the superconducting order parameter exists in the Au layer. For $t_{Au}\geq8$ ML, though two nodal planes are expected in the Au layer, the Nb layers are now in phase again. This enables a tunneling effect to occur between the Nb layers, and lifts the $\xi_\perp$ lock to $t_{Nb}$. In contrast to $\xi_\perp$, the parallel coherence length $\xi_{//}$ (not shown in figures) is independent of $t_{Au}$, and exhibits temperature dependence obeying an empirical expression $\xi_{//}\propto(1-T/T_c)^{-1}$, we have $\xi_{//}\sim40$ nm ($T=0.5T_c$) and $\xi_{//}\sim200$ nm ($T=0.9T_c$).

With regard to the oscillation period in $T_c(t_{Au})$, the Nb/Au/Nb trilayers and the previously studied Nb/Au/FM (FM=Fe, Co) trilayers exhibit different periods: the former have a period of ~1 nm [see Fig. 5(c)], while the latter a period of 2.1 nm [10, 11]. For a comprehensive understanding of these results, the boundary conditions for reflecting and/or scattering of quasiparticles at the Nb/Au and Au/FM boundaries should be evaluated properly. A difference in oscillation period by a factor of 2, however, provides a constraint on future theory.

**IV. DISCUSSIONS**

The analysis of the experimental data reasonably argued that a form of order-parameter oscillation is intrinsic to the proximity-induced Cooper pairs in the Au(111) layer. In other words, a pseudomagnetic effect on Cooper pairs holds in the Au layer. For the present, the mechanism underlying this order-parameter oscillation remains unclear. Any intrinsic paramagnetic moment observed in pure bulk Au [23] would be too small to account for the pseudomagnetic effect. The role of SOC in Au, however, merits further investigation. As mentioned in Sec. I, due to the broken inversion symmetry of fcc lattice in the direction perpendicular to the (111) plane, Rashba *type* SOC is induced in the whole Au(111) layer, in contrast to the well-studied Rashba SOC of surface origin [8]. We also add the fact that the Au(111) layer has a $C_{3v}$ point-group symmetry. Under this symmetry, we have to consider an in-plane structural inversion asymmetry—due to the threefold crystal symmetry around the <111> axis—that also contributes to the emergence of SOC as higher-order terms in the effective Hamiltonian [24]. In the argument above, a prominent contribution of the 5*d*-orbital to conduction electrons [6] should be called to mind.

Within the Au(111) layer, where SOC is induced because of the violation of three-dimensional (3D) inversion, the spin-singlet pairing and the spin-triplet pairing can be mixed by the proximity of a superconductor [25]. Here a parallel can be made with noncentrosymmetric superconductors such as $CePt_3Si$, in which, due to antisymmetric SOC, an FFLO-like state was expected to occur for the spin-triplet correlations even without magnetic field [26]. The high values of $J_c$ seen in Fig. 5(a) for our trilayers may support the presence of the spin-triplet correlations.

The value of $E_{ex}^{eff}$ (=84.6 meV) deduced from the fits to the $J_c(t_{Au})$ data (Fig. 6) is consistent with the SOC strength (120 meV) of conduction electrons in Au [7]. We infer that a pairing state should occur between two electrons on the SOC-induced split parts of the Fermi surface. One might refute this on the grounds that a difference of 84.6 meV in energy between the spin-split bands is too large for the electrons to make a pair, since the superconducting gap $\Delta_s$ is no more than 1.40 meV in Nb [27]. However this is not the case with regard to the proximity effects. As Kontos *et al.* [27] suggested for SC/FM junctions, even for exchange energies much higher—at least two orders of magnitude larger—than $\Delta_s$, superconducting correlations from SC still persist in FM, showing inhomogeneous superconductivity. The reason is that Cooper pairs are not instantaneously broken while they penetrate the area where pair breaking occurs [27]. We further get rid of another suspicion: SOC will not influence the proximity effect in the same way as the exchange field in FM, since SOC itself does not break time-reversal symmetry of the system. For this, we should point out that time-reversal symmetry is necessarily broken in the system by the presence of supercurrents [28-30].

The superconducting proximity effect in SOC materials has already been studied theoretically in some junction systems. In a multilevel quantum dot [28] and two-dimensional electron gas (2DEG) systems [29,31], Rashba and/or Dresselhaus SOC have a huge effect on the Josephson current. Especially for SC/(2DEG)/SC junctions with SOC in the 2DEG region, 0-π state transitions were found to occur as functions of the SOC strength and the 2DEG-layer length [29]. Clearly a theory is needed that treats the SOC effect on the proximity-induced Cooper pairs in a 3D SC/NM/SC system. The theoretical study of Nb/Au heterostructures (bilayer system), based on the self-consistent solution of the Kohn-Sham-Bogoliubov-deGennes equations, represents a first step in this direction [32]. Although SOC was not considered in these calculations, the nature of the Andreev bound states and the quasiparticle spectra were discussed in relation to the thickness of the Au layer. Importantly, the dependence of $T_c$ on the thickness of the Au layer in Nb/Au bilayers [10] was well reproduced by the calculations [33]. We anticipate a growing interest in material-specific calculations of this kind.

## V. CONCLUSION

We have studied a series of single-domain, epitaxially-grown Nb(110)/Au(111)/Nb(110) trilayers, for a range of thickness of the Au(111) layer, $0 < t_{Au} \leq 2.10$ nm. The critical current density $J_c$, superconducting coherence length $\xi_\perp$, and transition temperature $T_c$, all show a striking, non-monotonic dependence on $t_{Au}$. We compare this behavior with previous studies of superconducting multilayers incorporating ferromagnetic elements, and conclude that relative phase of the superconducting order parameter in the two Nb(110) layers changes sign for $t_{Au}$=4 ML(~1 nm) and 8 ML(~2 nm). For $4 \leq t_{Au} \leq 8$ ML, the trilayer therefore achieves a superconducting π junction. We tentatively ascribe these properties to spin-orbit coupling within the Au layer.

Our results suggest that the superconducting proximity effect in epitaxially grown Au offers a route to constructing π-state devices for application in superconducting electronics. This approach has the advantage that it eliminates the ferromagnetic elements typically found in π junctions based on superconducting multilayers, making possible a very high critical current density. Furthermore, if the modulation of the superconducting order parameter in the Au layer is due to spin-orbit coupling, it should be possible to achieve control the π junction at fixed $t_{Au}$, by applying an external electric field.

## ACKNOWLEDGEMENTS

The authors are pleased to acknowledge helpful conversations with James Annett, Gábor Csire, Bálazs Győrffy, and Bálazs Újfalussy. This work was supported by the Okinawa Institute of Science and Technology Graduate University (Japan), and the Engineering and Physical Sciences Research Council Grant No. EP/G049483/1 (United Kingdom).

**REFERENCES**

[1] A. I. Buzdin, Rev. Mod. Phys. **77**, 935 (2005).

[2] P. Fulde and R. A. Ferrell, Phys. Rev. **135**, A550 (1964).

[3] A. Larkin and Y. Ovchinnikov, Sov. Phys. JETP **20**, 762 (1965).

[4] A. I. Buzdin, L. N. Bulaevskii, and S. V. Panyukov, JETP Lett. **35**, 178 (1982).

[5] J. J. A. Baselmans, A. F. Morpurgo, B. J. van Wees, and T. M. Klapwijk, Nature (London) **397**, 43 (1999).

[6] A. Sekiyama, J. Yamaguchi, A. Higashiya, M. Obara, H. Sugiyama, M. Y. Kimura, S. Suga, S. Imada, I. A. Nekrasov, M. Yabashi, K. Tamasaku, and T. Ishikawa, New J. Phys. **12**, 043045 (2010).

[7] G.-M. Choi and D. Cahill, Phys. Rev. B **90**, 214432 (2014).

[8] S. LaShell, B. A. McDougall, and E. Jensen, Phys. Rev. Lett. **77**, 3419 (1996).

[9] J. Henk, M. Hoesch, J. Osterwalder, A. Ernst, and P. Bruno, J. Phys.: Condens. Matter **16**, 7581 (2004).

[10] H. Yamazaki, N. Shannon, and H. Takagi, Phys. Rev. B **73**, 094507 (2006).

[11] H. Yamazaki, N. Shannon, and H. Takagi, Phys. Rev. B **81**, 094503 (2010).

[12] C. Sürgers and H. v. Löhneysen, Appl. Phys. A **54**, 350 (1992).

[13] See, for example, J. F. Annett, *Superconductivity, Superfluids and Condensates* (Oxford University Press, Oxford, 2004) p.81.

[14] E. M. Gyorgy, R. B. van Dover, K. A. Jackson, L. F. Schneemeyer, and J. V. Waszczak, Appl. Phys. Lett. **55**, 283 (1989).

[15] A. Pan, M. Ziese, R. Höhne, P. Esquinazi, S. Knappe, and H. Koch, Physica C **301**, 72 (1998).

[16] A. Yu. Rusanov, M. B. S. Hesselberth, and J. Aarts, Phys. Rev. B **70**, 024510 (2004).

[17] G. Carneiro, Phys. Rev. B **57**, 6077 (1998).

[18] G. Stejic, A. Gurevich, E. Kadyrov, D. Christen, R. Joynt, and D. C. Larbalestier, Phys. Rev. B **49**, 1274 (1994).

[19] Y. Mawatari, and K. Yamafuji, Physica C **228**, 336 (1994).

[20] Y. Blum, A. Tsukernik, M. Karpovski, and A. Palevski, Phys. Rev. Lett. **89**, 187004 (2002).

[21] See, for example, C. Kittel, *Introduction to Solid State Physics* 6th edition (John Wiley & Sons, New York, 1986) p.134.

[22] V. Shelukhin, A. Tsukernik, M. Karpovski, Y. Blum, K. B. Efetov, A. F. Volkov, T. Champel, M. Eschrig, T. Löfwander, G. Schön, and A. Palevski, Phys. Rev. B **73**, 174506 (2006).

[23] M. Suzuki, N. Kawamura, H. Miyagawa, J. S. Garitaonandia, Y. Yamamoto, and H. Hori, Phys. Rev. Lett. **108**, 047201 (2012).

[24] Sz. Vajna, E. Simon, A. Szilva, K. Palotas, B. Ujfalussy, and L. Szunyogh, Phys. Rev. B **85**, 075404 (2012).

[25] L. P. Gor'kov and E. I. Rashba, Phys. Rev. Lett. **87**, 037004 (2001).

[26] H. Tanaka, H. Kaneyasu, and Y. Hasegawa, J. Phys. Soc. Jpn. **76**, 024715 (2007).

[27] T. Kontos, M. Aprili, J. Lesueur, and X. Grison, Phys. Rev. Lett. **86**, 304 (2001).

[28] L. Dell'Anna, A. Zazunov, R. Egger, and T. Martin, Phys. Rev. B **75**, 085305 (2007).

[29] Z. H. Yang, Y. H. Yang, J. Wang, and K. S. Chan, J. Appl. Phys. **103**, 103905 (2008).

[30] O. V. Dimitrova and M. V. Feigel'man, J. Exp. Theor. Phys. **102**, 652 (2006).

[31] X. Liu, J. K. Jain, and C.-X. Liu, Phys. Rev, Lett. **113**, 227002 (2014).

[32] G. Csire, B. Újfalussy, J. Cserti, and B. Györffy, Phys. Rev. B **91**, 165142 (2015).

[33] G. Csire, J. Cserti, I. Tütto, and B. Újfalussy, arXiv: 1601.07038.

| cap Au(111) | 1.74 nm |
| Nb(110) | 20.0 nm |
| Au(111) | $t_{Au}$ |
| Nb(110) | 20.0 nm |
| $Al_2O_3(11\bar{2}0)$ | |

**FIG. 1.** Schematic diagram of a vertical section of the sample and layer thicknesses.

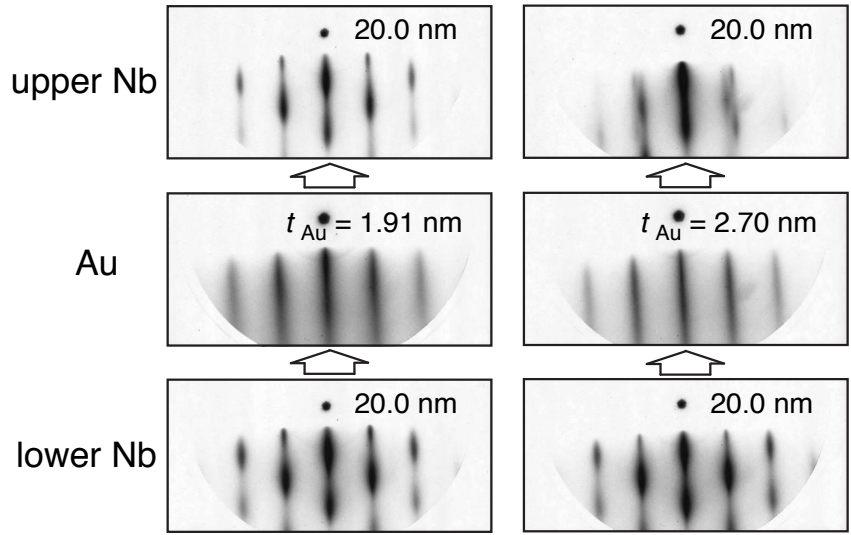

**FIG. 2.** Reversal images of RHEED patterns obtained in the growth process of the Nb/Au[$t_{Au}$]/Nb trilayer for $t_{Au}$=1.91 nm (left column; typical of the samples for 0.40≤$t_{Au}$≤2.10 nm) and $t_{Au}$=2.70 nm (right column; typical of the samples for $t_{Au}$≥2.25 nm). The direction of the incident electron beam is parallel to <1$\bar{1}$0> of the Nb(110) layer.

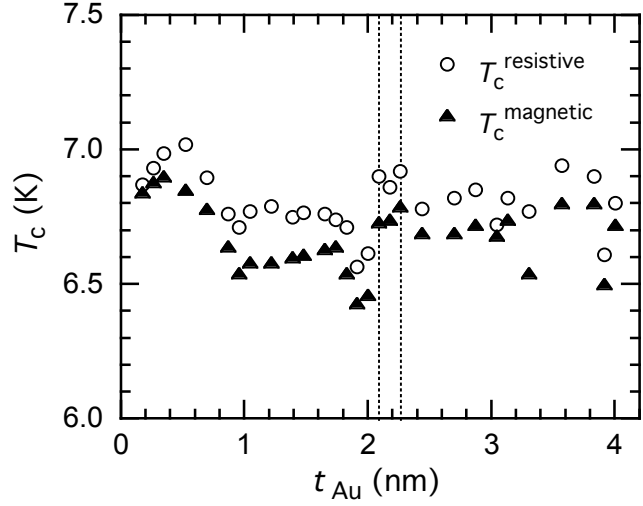

**FIG. 3.** Superconducting transition temperature $T_c$: $T_c^{\text{resistive}}$ and $T_c^{\text{magnetic}}$ as a function of $t_{\text{Au}}$, where $T_c^{\text{resistive}}$ was defined as the temperature at 50% of the residual resistivity and $T_c^{\text{magnetic}}$ was as the onset point of the diamagnetic transition. The error of $T_c$ is within each symbol. The broken lines are drawn at 2.10 and 2.25 nm, indicating a region of structural transition for the upper Nb layer.

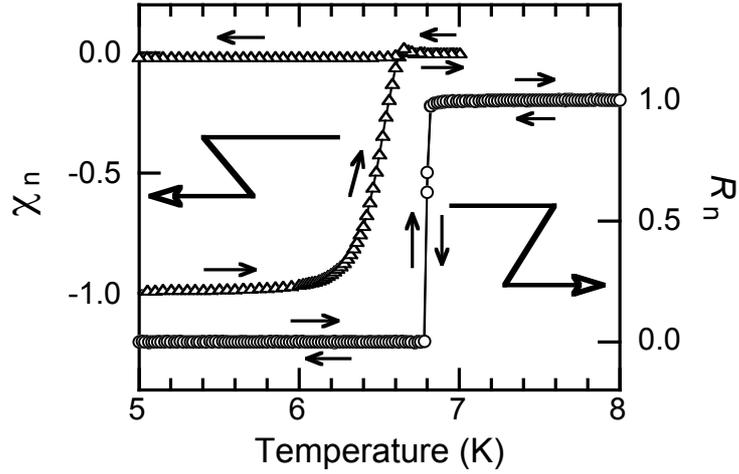

**FIG. 4.** Typical temperature dependences of normalized magnetic susceptibility $\chi_n$ ($H$=0.1 Oe, $H \perp$ surface) and of normalized resistivity $R_n$ ($H$=0 Oe) for the $t_{Au}$=1.48 nm sample. A transition width of 0.04 K (10-90% criterion) is seen for $R_n$. Magnetic susceptibility measurements were performed after zero-field cooling, using a superconducting quantum interference device magnetometer (Quantum Design MPMSXL). Resistivity measurements were carried out in a standard four-terminal configuration using low-frequency ac technique.

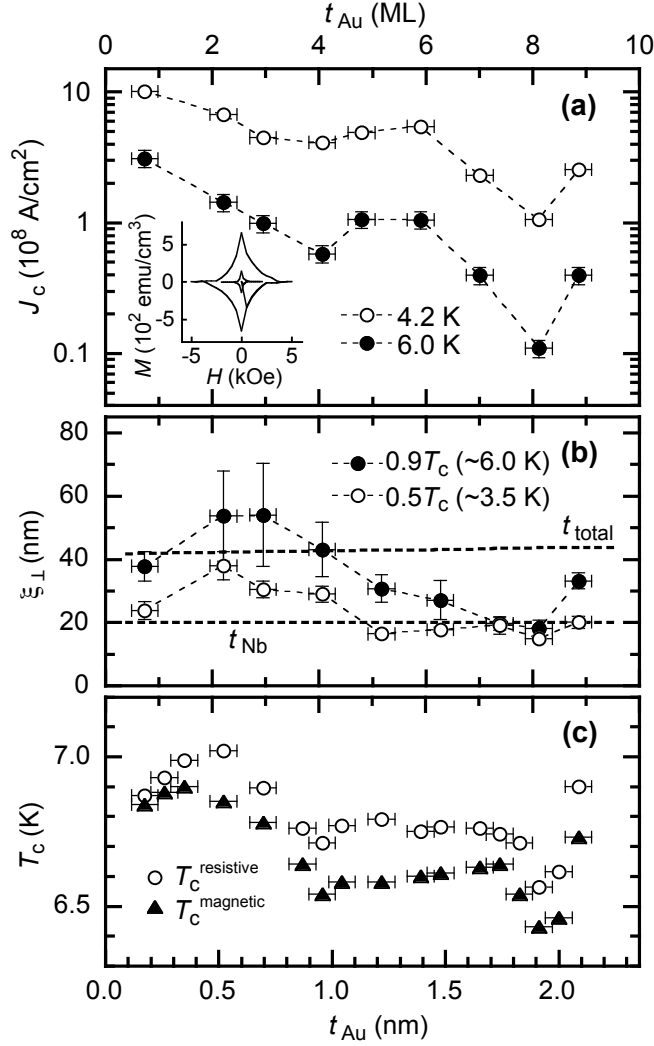

**FIG. 5.** Superconducting properties as functions of $t_{Au}$ for the Nb/Au[$t_{Au}$]/Nb trilayers: (a) critical current density $J_c$ at 0 Oe , (b) perpendicular coherence length $\xi_\perp$ (the vertical error bars are due to the uncertainty in $H_{c2\perp}$ and $H_{c2//}$), and (c) transition temperature $T_c$ (part of Fig. 3). One atomic monolayer (ML) corresponds to $t_{Au}$=0.2355 nm. In (b), the horizontal $t_{Nb}$ line ($\xi_\perp$=20.0 nm) corresponds to the thickness of a single Nb layer; the inclined $t_{total}$ line shows a thickness of the whole sample (excluding the substrate): $t_{total}$=20.0+$t_{Au}$+20.0+1.74 nm. Inset in (a): Typical $M$-$H$ curves ($H$ // sample plane) measured at 4.2 K (outer) and 6.0 K (inner) for the $t_{Au}$=1.13 nm (~5 ML) sample. Diamagnetic contribution from the substrate has been subtracted.

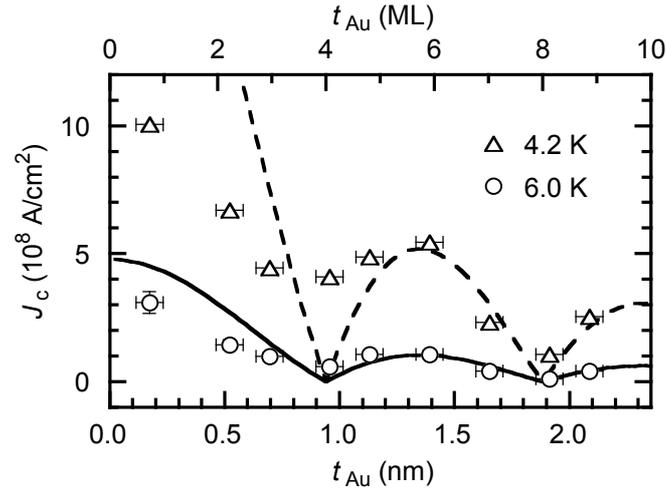

**FIG. 6.** Critical current density $J_c$ of the Nb/Au[$t_{Au}$]/Nb trilayers at 4.2 K and 6.0 K. Data look exactly like FM/SC multilayers sketched in Ref. [20]. Following Refs. [1] and [20] (using Eq. (2)), solid and dashed lines show theoretical fits to the data for 6.0 K and 4.2 K, respectively. Fits are obtained as a function of $E_{ex}^{eff}$ (=84.6 meV), for $v_f^{Au}$=1.39×10$^6$ m/sec and the ratio $J_c^0$(4.2 K)/$J_c^0$(6.0 K)=5.0.